# Renal Parenchymal Area and Kidney Collagen Content

Jake A. Nieto[1], Janice Zhu[1], Bin Duan, Jingsong Li, Ping Zhou, Latha Paka, Michael A. Yamin, Itzhak D. Goldberg and Prakash Narayan*

1, Equal contribution

*, Corresponding Author

Address: Angion Biomedica Corp., 51 Charles Lindbergh Blvd, Uniondale, 11553, New York, USA

Ph: 516 326 1200

Fax: 516 222 1359

Email: pnarayan@angion.com

**Abstract**

The extent of renal scarring in chronic kidney disease (CKD) can only be ascertained by highly invasive, painful and sometimes risky tissue biopsy. Interestingly, CKD-related abnormalities in kidney size can often be visualized using ultrasound. Nevertheless, not only does the ellipsoid formula used today underestimate true renal size but also the relation governing renal size and collagen content remains unclear. We used coronal kidney sections from healthy mice and mice with renal disease to develop a new technique for estimating the renal parenchymal area. While treating the kidney as an ellipse with the major axis the polar distance, this technique involves extending the minor axis into the renal pelvis. The calculated renal parenchymal area is remarkably similar to the measured area. Biochemically determined kidney collagen content revealed a strong and positive correlation with the calculated renal parenchymal area. The extent of renal scarring, i.e. kidney collagen content, can now be computed by making just two renal axial measurements which can easily be accomplished via noninvasive imaging of this organ.

## Introduction

Given the prevalence of diabetes, hypertension and Metabolic Syndrome, chronic kidney disease (CKD) is reaching epidemic proportions across the world [1,2]. Characterized by scarring or accumulation of fibrillar collagen within the renal interstitium, CKD is associated with a progressive decline in glomerular filtration rate (GFR) reflected by rising serum creatinine (SCr). However, existing disease is often diagnosed late because clinically meaningful changes in SCr occur long after substantial and irreversible scar formation [3,4]. Further compounding both disease diagnosis and prognosis is the fact that highly invasive renal biopsy remains the mainstay for determining the extent of renal scarring [5]. Interestingly, fibrosis-related abnormalities in kidney dimension can be visualized by noninvasive sonography [6-8]. Renal length or major axis, renal width or minor axis and renal thickness measurements can be incorporated into an ellipsoid formula to yield kidney size [9-11]. Unfortunately, there is insufficient information relating renal dimension with the amount of tissue interstitial collagen. Second, the standard ellipsoid formula underestimates true kidney size confounding any inferences of tissue collagen content [9-11].

In the present study, we used a mouse model of CKD to develop a modified elliptical formula that better represents true renal parenchymal area. We then formulated a relationship between calculated renal parenchymal area and total tissue collagen.

**Methods**

**Animal Model:** The study protocol, designed to induce renal fibrosis in mice, was submitted to and approved by the Angion Biomedica Corp. Institutional Animal Care and Use and Committee. Animals were allowed to acclimatize for a minimum of 5 days prior to use and had free access to water and standard rodent chow. Adult male CD-1 mice (~30-35 g) were anesthetized with ketamine (25 mg/kg, intraperitoneal) and xylazine (5 mg/kg, ip) and placed on a heating pad table to maintain ~37.5°C core body temperature. A midline laparotomy was made and the right kidney removed was removed. Extended release buprenorphine (0.65 mg/kg, subcutaneous) was administered prior to returning animals to their cages. One week later, animals were placed on 1% NaCl (drinking water) and deoxycorticosterone acetate (DOCA, 1 mg/kg, subcutaneous) injected twice weekly for the first 3 weeks [12,13]. Eight weeks after nephrectomy, animals were sacrificed and the left kidney retrieved. Age-matched, surgery-and DOCA-naive animals on regular drinking water were used as the baseline control. Left kidneys from these animals were retrieved at sacrifice. Kidneys were weighed and sliced coronally under a dissecting microscope (4X). One half of the kidney was placed in 10% formalin for subsequent sectioning (coronal sections 5 µm apart) and staining with hematoxylin & eosin (H&E). The other half of the kidney was weighed and submitted to hydroxyproline analysis using a previously described method [14]. Total kidney hydroxyproline values were converted to total kidney collagen (µg/kidney) content [14].

**Renal Parenchymal Area:** H&E-stained renal sections were photographed (Nikon) and analyzed using NIS-Elements D 3.1 software by an observer blinded to the collagen content of that kidney. Images were superimposed on a precalibrated grid and the major (a), minor (b) and the extended minor ($b_e$) axes (mm) measured (see Figs 1 and 3). Renal parenchymal area (mm$^2$) was measured using the "area measurement" tool available in the software and also calculated from equations 1 and 2.

**Data Analysis**: Both collagen and the corresponding renal parenchymal area measurement were obtained from a total of 30 kidneys, 10 from the healthy cohort and 20 from the diseased cohort. Microsoft Excel 2010 curve fitting software was used to generate all scatterplots. Since a linear relation was observed between the 2 variables in each of the scatterplots, both Pearson product moment (r) and Spearman's rho ($r_s$) were calculated from the trend line. To determine whether the relationship between the 2 variables was significant, r or $r_s$ and the sample size (n=30) were entered into an online calculator [15]. A $p < 0.05$ was considered to be statistically significant.

**Results**

Shown in Fig 1 is an H&E-stained renal coronal section from a uninephrectomized mouse administered DOCA and 1% NaCl (drinking water) with the major (a) and minor (b) axes delineated.

Fig 1

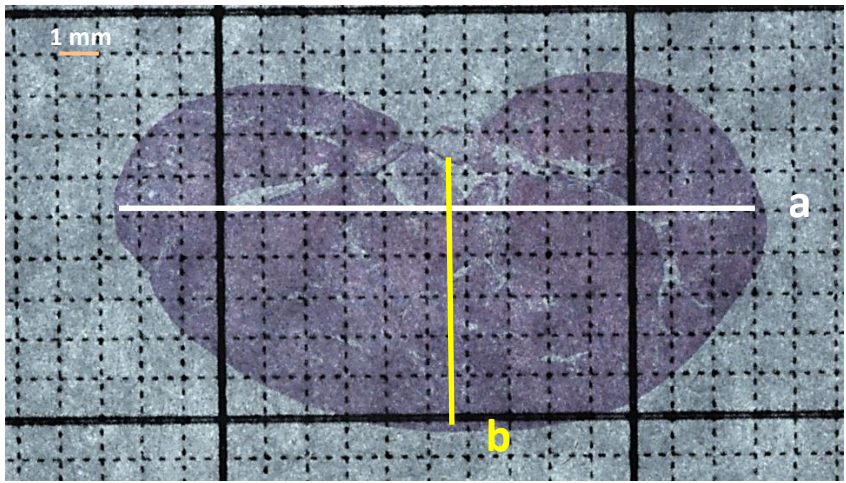

**Fig 1: Renal Parenchyma Area.** An H&E-stained coronal section (4X) from the left kidney of a uninephrectomized mouse administered DOCA and NaCl. The section has been superimposed on a 1 mm$^2$ grid. The white bar represents renal length or the major axis (a) whereas the yellow bar represents renal width or the minor axis (b). The renal parenchymal area can be measured using a precalibrated measuring tool or calculated from a and b.

The equation for the area of an ellipse, viz.,

$$A = (\pi * a * b) / 4 \qquad (1)$$

was used to calculate the renal parenchyma area (*A*) and correlate it with the measured parenchymal area ($A_m$). As seen in Fig 2A, there is a very high correlation between these two variables (r = 0.9, p < 0.01; $r_s$ = 0.89, p < 0.01) [16]. Nevertheless, consistent with the published literature [9-11], use of the standard elliptical formula underestimates true renal dimension as *A* was only 86 ± 1% of $A_m$ (p < 0.01; Fig 2B).

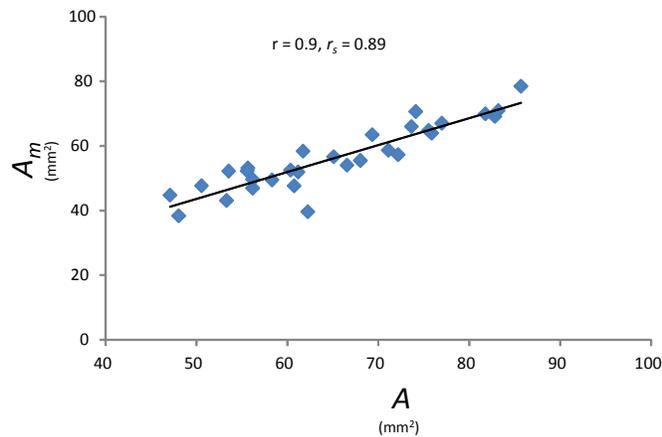

Fig 2A

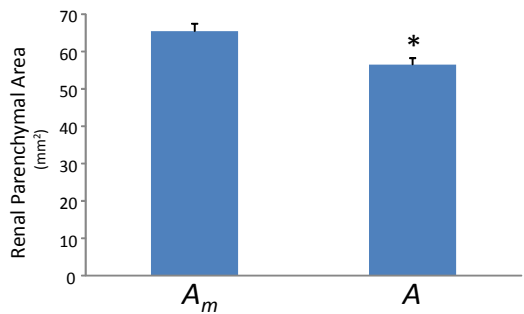

Fig 2B

**Fig 2: Measured ($A_m$) vs. Calculated (*A*) Renal Parenchymal Areas.** (**A**) A scatter plot of $A_m$, the measured renal parenchymal area vs. *A*, the renal parenchymal area calculated by entering the axial dimensions, length (a) and width (b), into the elliptical formula. The correlation was significant (p < 0.01). (**B**) The calculated parenchymal area from 30 kidneys is only 86% of the measured parenchymal area. This difference is significant (*. p < 0.01).

To obtain greater fidelity toward $A_m$, the minor axis was extended into the renal pelvis as shown in Fig 3.

Fig 3

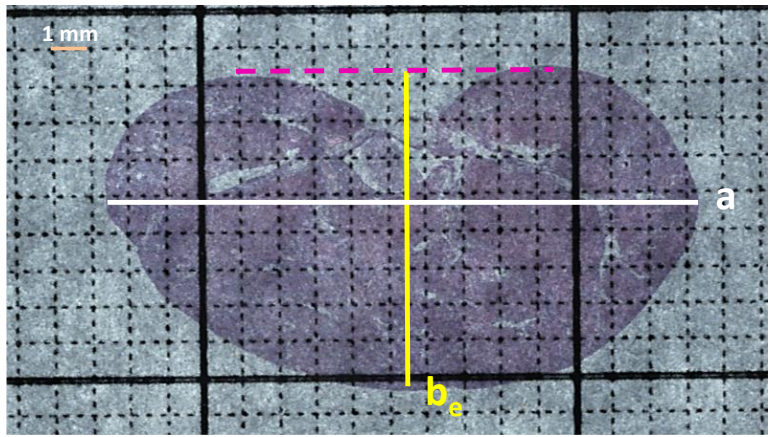

**Fig 3: Renal Parenchyma Area.** An H&E-stained coronal section (4X) from the left kidney of a uninephrectomized mouse administered DOCA and NaCl. The section has been superimposed on a 1 mm² grid. The white bar represents renal length or the major axis (a) whereas the yellow bar represents the extended minor axis ($b_e$), which has been extended into the renal pelvis until it intersects the pink dashed bar. The renal parenchymal area can be measured using a precalibrated measuring tool or calculated from a and $b_e$.

A modified elliptical area equation, viz.

$$A_e = (\pi * a * b_e) / 4 \qquad (2)$$

was used to recalculate renal parenchymal areas ($A_e$) and correlate it with $A_m$. As seen in Fig 4A, use of an extended minor axis in the elliptical equation returned calculated areas that correlated very highly with the measured areas (r = 0.97, p < 0.01; $r_s$ = 0.96, p < 0.01) [16]. Importantly, for the sample set, $A_e$ was 100.9 ± 0.7% of $A_m$ (p not significant; Fig 4B).

Fig 4A

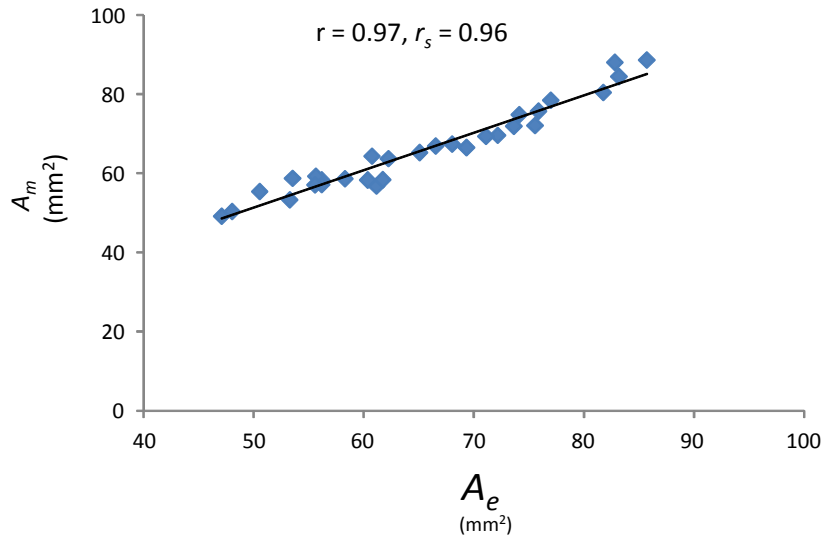

Fig 4B

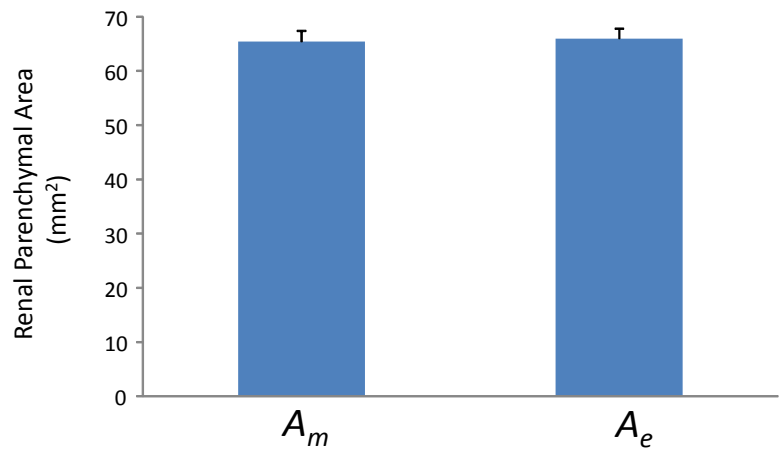

**Figure 4: Measured ($A_m$) vs. Calculated ($A_e$) Renal Parenchymal Areas.** (**A**) A scatter plot of $A_m$, the measured renal parenchymal area vs. $A_e$, the renal parenchymal area calculated by entering the axial dimensions of length (a) and modified width ($b_e$) into the elliptical formula. The correlation was significant (p < 0.01). (**B**) The calculated average parenchymal area from 30 kidneys is not different from the measured areas.

Collagen content of these kidneys, measured using the hydroxyproline assay and using a conversion factor of 13.5, were plotted against both $A_m$ as well as $A_e$. As seen in Fig 5, a high correlation was observed between kidney collagen and $A_m$. (r = 0.8, p < 0.01, $r_s$ = 0.79, p < 0.01) [16].

Fig 5

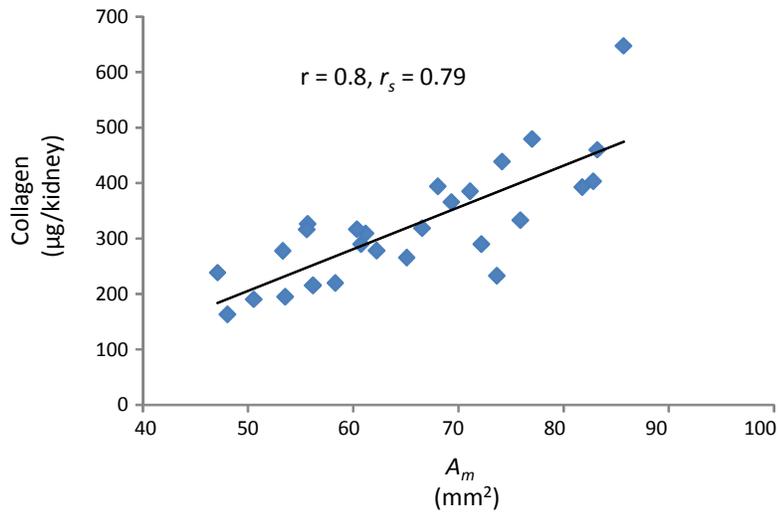

**Figure 5: Kidney Collagen vs. Measured Renal Parenchymal Area.** Collagen content from healthy and diseased kidneys was correlated with the corresponding measured renal parenchymal areas ($A_m$). The correlation was significant.

As seen in Fig 6, plotting kidney collagen content vs. $A_e$ also yielded a high correlation with r = 0.8, p < 0.01, $r_s$ = 0.77 and p < 0.01 [16]. In fact, the correlation with kidney collagen for $A_m$ and $A_e$ were very similar, once again suggesting that $A_e$ can be substituted for $A_m$.

Fig 6

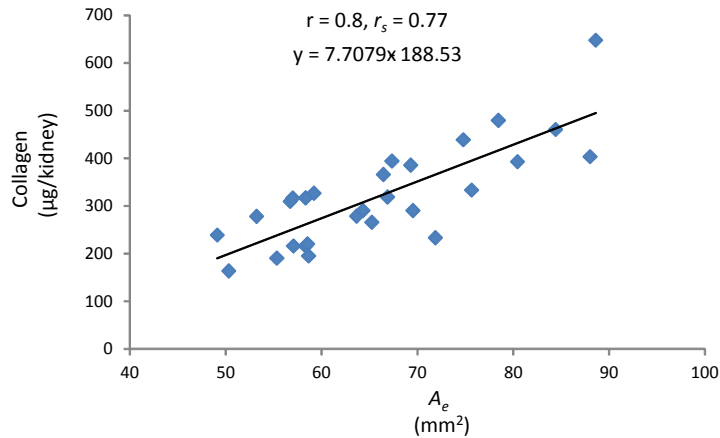

**Figure 6: Kidney Collagen vs. Calculated Renal Parenchymal Area.** Collagen content from healthy and diseased kidneys was correlated with the corresponding renal parenchymal areas ($A_e$) calculated using an elliptical equation with an extended minor axis ($b_e$). The correlation was significant. The equation inset describes the relation between kidney collagen content and $A_e$.

The relation between kidney collagen content and renal parenchymal area can be described by the following formula

$$\text{Collagen (µg/kidney)} = [7.7 * (\pi * a * b_e / 4) - 188.5] \qquad (3)$$

**Discussion**

Using a murine model of CKD, we have developed a new method involving just two linear, axial measurements that allows for calculation of renal parenchymal area. The calculated parenchymal area is in excellent agreement with the measured or true renal parenchymal area. Furthermore, we have now defined the relationship between renal parenchymal area and total collagen content within the kidney enabling relatively easy and rapid determination of the extent of kidney fibrosis.

Management of the CKD patient continues to present a challenge for the nephrologist given that changes in SCr are long preceded by extracellular matrix accumulation within the renal interstitium, and given that highly invasive and painful tissue biopsy remains the mainstay for determining the extent of scarring [5]. Furthermore, many patients at highest risk for CKD, including those with coagulation disorders and uncontrolled hypertension, are often not candidates for biopsy [17]. A noninvasive method to capture and track kidney collagen content can obviate the need for repeated biopsies. Intriguingly parenchymal echogenicity and abnormalities in renal size are often evident in CKD [6-8]. While measurement of renal dimensions and calculation of renal volume based on the ellipsoid formula is now standard practice [10], there are little, if any, data translating this information to kidney collagen content. Furthermore, a number of clinical studies [9-11] has shown that use of this formula consistently underestimates kidney volume reported at autopsy.

In the present study, we investigated the renal parenchymal area and corresponding kidney collagen content from healthy mice and DOCA-salt-uninephectomized mice, a standard and well-characterized model of murine CKD [12-13]. A salient feature of this model is that, unlike the subtotal nephrectomy model, the kidney is not surgically perturbed or partially ablated therefore lending itself to reliable measurements of length and width. Consistent with the afore-referenced reports, entering the standard axial (length and width) measurements into an elliptical formula underestimated renal parenchymal area. By contrast, use of a minor axis measurement that includes the renal pelvis in the elliptical formula yields a parenchymal area that is remarkably consistent with the measured area.

The other hallmark finding of this study was an understanding of the relation governing renal parenchymal area with kidney collagen content. In our sample set, comprising healthy and diseased kidneys, tissue collagen spanned a 4-fold dynamic range. Over this range, renal parenchymal area tracked tissue collagen evidenced by very similar Pearson product moment and Spearman Rho values, with the latter speaking not only to the strength but also the direction of this relation. Importantly, both the measured parenchymal area, and the parenchymal area calculated using an extended minor axis, returned similar r and $r_s$ values vis. a vis. tissue collagen content. This finding is of translational interest in that, as described by equation 3, kidney collagen content can now be computed from just two axial measurements across the kidney viz. the major axis and a minor axis that extends into the renal pelvis.

There are certain limitations to our findings. The present study sampled kidneys from a specific model of murine CKD at single timepoint of 8 weeks. It remains to be determined whether our findings hold true across different timepoints within this model or in other models of CKD such as diabetic nephropathy or in kidneys from higher species including humans. The effects of reverse renal remodeling and potential hysteresis in response to interstitial tissue catabolism on renal parenchymal area remain to be determined. Finally, our findings were drawn from H&E-stained coronal renal slices and may not fully translate to findings drawn using non-invasive imaging modalities such as ultrasound.

These limitations notwithstanding, our results form the foundation for developing a calculator for fibrosis from measurements made during noninvasive renal imaging. Such a calculator will find clinical use along the lines of other existing calculators for renal and liver diseases such as the modified diet in renal disease (MDRD), CKD-epidemiology collaboration (CKD-EPI), FIB-4 and aspartate aminotransferase-to-platelet ratio index (APRI) calculators [18-20]. Such a calculator will not only represent a patient-friendly and relatively inexpensive method to track disease progress and aid in the management of this population but can also potentially be used in clinical trials of drugs that work by reducing the deposition of tissue collagen.

**Acknowledgements**

This study was funded, in part, by awards 2R44DK085771-02A1 (NIDDK) and W81XWH-14-1-0411 (DoD) made to PN and Angion Biomedica Corp., NY.